%% file: feng.tex
\RequirePackage[hyphens]{url}
\documentclass[12pt]{spieman}  
\pdfoutput=1
\usepackage{amsmath,amsfonts,amssymb}
\usepackage{graphicx}
\usepackage{setspace}
\usepackage{tocloft}

\usepackage[per-mode=symbol]{siunitx}
\DeclareSIUnit{\sample}{sample}
\DeclareSIUnit{\flop}{flop}
\usepackage{dcolumn}
\usepackage{pgfplotstable}
\pgfplotsset{compat=1.17}
\pgfkeys{/pgf/number format/set thousands separator={\,}}
\usepackage{booktabs}
\usepackage{underscore}
\usepackage{lineno}

\title{Efficient channelization on a Graphics Processing Unit}

\author[a,*]{Bruce Merry}
\affil[a]{South African Radio Astronomy Observatory, 2 Fir Street, Cape Town,
South Africa, 7925}

\cftpagenumbersoff{figure}
\cftpagenumbersoff{table}
\begin{document}
\maketitle

\begin{abstract}
  We present an implementation of a channelizer (F-engine) running on a
  Graphics Processing Unit (GPU). While not the first GPU implementation of a
  channelizer, we have put significant effort into optimizing the
  implementation. We are able to process four antennas each with
  \qty{2}{\giga\sample\per\second}, 10-bit dual-polarized input and
  8-bit output, on a single commodity GPU. This fully utilizes the available
  PCIe bandwidth of the GPU. The system is not as optimized for a single
  high-bandwidth antenna, but handles
  \qty{6.2}{\giga\sample\per\second}, limited by single-core CPU performance.
\end{abstract}


{\noindent \footnotesize\textbf{*}Bruce Merry,  \linkable{bmerry@sarao.ac.za} }


\input{intro.tex}
\input{background.tex}
\input{implementation.tex}
\input{unzip.tex}
\input{results.tex}
\input{conclusions.tex}

\subsection*{Acknowledgments}
The MeerKAT telescope is operated by the South African Radio Astronomy
Observatory, which is a facility of the National Research Foundation, an
agency of the Department of Science and Innovation.

Boston Limited and Boston IT Solutions South Africa provided access to test
systems that were used in the development of the software described in this paper.

\subsection*{Data, Materials, and Code Availability}
The channelizer implementation described in this paper is available under a
BSD license at \url{https://github.com/ska-sa/katgpucbf}.

\bibliography{feng.bib}
\bibliographystyle{spiejour}   

\vspace{2ex}\noindent\textbf{Bruce Merry} is a senior developer at the South
African Radio Astronomy Observatory. He received his BSc (Hons) and PhD
degrees in computer science from the University of Cape Town in 2003 and 2007,
respectively. He has published papers in computer graphics, accelerated
computing, and radio astronomy software. His specialities include
high-performance networking and GPU computing.

\listoffigures
\listoftables

\end{document}

%% file: intro.tex
\section{Introduction}
\label{sect:intro}
The MeerKAT radio telescope has an existing correlator-beamformer based on Field
Programmable Gate Arrays (FPGAs), which has been previously described
\cite{meerkat-cbf}. The MeerKAT Extension project is currently underway to
add more dishes with longer baselines\cite{mke}. Since the MeerKAT correlator depends on
a number of hardware components that have reached end-of-life (particularly
the Hybrid Memory Cube memory) and there were concerns that the design would
not scale up, a new correlator is being designed rather than
expanding the existing correlator.

The FPGA development process for MeerKAT was plagued by long compile times
(usually overnight), difficult-to-use tools, and rigid designs: each channel count
used a different design, and changes had to be manually copied between
designs. While FPGA development tools have since improved, it nevertheless
remains challenging to achieve high performance \cite{fgpa-vs-gpu}.

Graphics Processing Units (GPUs) offer an alternative with a mature ecosystem
and a more convenient development process. They have been used for some years
for the correlation (X) step in F-X correlators\cite{xgpu}, but the only GPU-based
channelizer (F step) of which we are aware is the Cobalt correlator (used by
LOFAR) \cite{cobalt,accel-compare}. There are also GPU-based spectrometers
such as at the Green Bank Telescope \cite{gbt-spectrometer} and
Atacama Compact Array \cite{kasi-spectrometer}, but these do not include the
delay correction needed for an F-X correlator.

We first developed a proof-of-concept
channelizer which implemented the data-path functionality of the MeerKAT
wide-band correlator. Partly based on the good results from this
proof-of-concept, we elected to pursue a fully GPU-based correlator for the
MeerKAT Extension. This paper describes the implementation and tuning of our
GPU-based channelizer.

Section~\ref{sect:background} describes the functionality included in our
channelizer, and summarizes the programming model for GPUs.
Section~\ref{sect:implementation} details the initial software
implementation. We then describe a significant optimization in
Sec.~\ref{sect:unzip}. We finish with results (Sec.~\ref{sect:results}) and
conclusions (Sec.~\ref{sect:conclusions}).

%% file: background.tex
\section{Background}
\label{sect:background}

\subsection{Channelization}
The exact steps performed by a channelizer are likely to vary from one
instrument to the next. The MeerKAT wide-band channelizers (both the original
FPGA design and our new GPU design) perform the following steps. These are
essentially the same as those performed by Cobalt \cite{cobalt}:

\begin{enumerate}
  \item
    Digitized samples are received. The MeerKAT digitizers produce 10-bit
    signed integer voltage samples. The data rate varies
    depending on the observing band, but is up to
    \qty{1750}{\mega\sample\per\second} (\qty{17.5}{\giga\bit\per\second})
    for each of the two polarizations of each antenna.
  \item
    A coarse delay is applied. Signals of interest will arrive at
    different antennas at different times due to the finite speed of light.
    Before correlating them, this delay must be corrected. Coarse delay is
    applied in the time domain and operates only on a whole number of samples.
  \item
    A polyphase filter bank (PFB)\cite{spectrometers} is applied to convert
    time-domain data to the frequency domain. A PFB is essentially a set of
    strided finite impulse response (FIR) filters followed by a Fourier
    transform. Suppose we wish to compute a spectrum with start time $t_0$
    with $n$ channels. The PFB has a set of weights $w_{i,j}$ ($0 \le i < 2n$,
    $0 \le j < T$ where $T$ is the number of ``taps'').  The filtered sample
    $g_t$ at time $t_0 + t$ ($0 \le t < 2n$, where time is measured in
    samples) is
    \begin{equation}
      g_t = \sum_{j=0}^{T-1} s_{t_0 + t - c + 2nj} \cdot w_{t, j},
      \label{eq:fir}
    \end{equation}
    where $s$ contains the original samples and $c$ is the coarse delay.
    The filtered samples $g_t$ are then put
    through a $2n$-element real-to-complex Fourier transform, from which only
    the $n$ non-negative frequencies are retained. The Nyquist frequency
    (which lacks phase information), is dropped so that the result is a
    convenient power-of-two size.

    MeerKAT supports \num{1024}, \num{4096} or \num{32768} frequency channels,
    and the MeerKAT Extension will additionally
    support \num{8192} channels. The MeerKAT PFBs use up to 16 taps in the FIR
    filters to improve the channel isolation.
  \item
    A fine delay is applied. This is the residual delay not corrected by the
    coarse delay step, and is corrected by phase rotation in the frequency
    domain.
  \item
    Bandpass correction is applied: each value is multiplied by a complex,
    channel-dependent correction factor.
  \item
    The internal representation is quantized to 8-bit signed Gaussian
    integers, arranged into packets and transmitted to the network.
\end{enumerate}
In the signal processing steps described above, the two polarizations remain
independent of each other. However, to maintain compatibility with the MeerKAT
packet formats, each output packet contains data from both polarizations, and
hence the channelizer needs to operate on both polarizations together. Doing
so also allows for new features in the future, such as correcting for
polarization leakage. We refer to a
pipeline performing all the steps above for two polarizations of a single
antenna as an F-engine. A single server may run multiple F-engines as
independent processes.

\subsection{Network Protocol}
MeerKAT uses the Streaming Protocol for Exchange of Astronomical
Data (SPEAD)\cite{spead}, deployed over multicast UDP. This is a
protocol for transmitting multi-dimensional arrays of data with
associated metadata (such as timestamps). The basic protocol data unit is the
\emph{heap}, which may be fragmented into multiple UDP packets and reassembled
by the receiver.

Digitizers send voltage samples in \num{4096}-sample heaps, each comprising a
single packet. The 10-bit samples are packed, so the heaps contain \num{5120}
bytes of payload. The two polarizations are sent independently.

To reduce the number of F-engine output heaps, each heap contains data for 256 spectra.
The data in a heap is a $c \times 256 \times 2$ array of Gaussian integers,
where $c$ is the number of channels sent to each consumer, and for
MeerKAT can be anywhere from 4 to \num{2048}. Output heaps may comprise
multiple UDP packets, as they are typically larger than the largest possible
UDP packet.

\subsection{Graphics Processing Units}
Our target GPUs are those from NVIDIA, and so we will use the terminology used
by CUDA (NVIDIA's programming toolkit). GPUs from other vendors are similar
but use different terminology. CUDA-capable GPUs have multiple levels of
parallelism:

\begin{description}
  \item[Threads] are the finest level, and are conceptually similar to CPU
    threads. Each thread has its own registers. Threads are programmed as if
    they have independent control flow, but in practice there are limitations
    to this, and dynamic control flow at the thread level can reduce
    performance.
  \item[Warps] are groups of 32 threads, and are the unit of scheduling. For
    best performance, all the threads in a warp execute the same instruction
    at the same time, and access adjacent memory locations.
  \item[Thread blocks] are sets of threads that execute concurrently on a
    single Streaming Multiprocessor (SM)---one of the hardware units of the
    GPU. Threads in the same block can communicate through a high-speed shared
    memory that is local to the SM. Each SM also has an L1 cache.
  \item[Grids] are the coarsest level. The CPU dispatches work to the GPU as a
    grid of thread blocks. Every thread executes the same program, but is
    assigned a unique index that allows it to be differentiated from other
    threads. A grid may contain more thread blocks than the GPU has the
    resources to handle concurrently, in which case some thread blocks may
    only start after others have completed.
\end{description}

The GPUs we have tested all connect to a host system via 16 lanes of a PCIe
4.0 bus. The CUDA busGrind tool typically shows that NVIDIA GPUs can
sustain \qty{26}{\giga\byte\per\second} for unidirectional traffic, and
\qty{21}{\giga\byte\per\second} each way for bidirectional traffic. This is
1--2 orders of magnitude less than the bandwidth of the RAM on the GPU and can
easily become a bottleneck for data streaming applications.


%% file: implementation.tex
\section{Implementation}
\label{sect:implementation}
Unlike other correlators of which we are aware, the CPU parts of our
correlator are implemented entirely in Python, while the GPU kernels are written in
CUDA C++. While not generally known for its performance, Python's high-level
nature has lead to high developer productivity. All the compute-intensive work
is either performed on the GPU or handled by off-the-shelf libraries such as
spead2 or numpy that use C/C++ internally. Some code is carefully
written to ensure that Python does not get used for performance-critical inner
loops.

\subsection{Batching}
The programming model in CUDA (and other GPU programming interfaces) is
batch-oriented rather than based on a continuous stream of data. Large
batches of work (millions of threads) allow for sufficient parallelism to keep
the GPU fully utilized. We thus break the input stream into \emph{chunks} of a
few million samples. The output stream is similarly decomposed into chunks,
but for reasons we will explain later, they are not in one-to-one
correspondence. Because the Python code is involved at the batch level (rather
than on individual samples or packets), large batches also help amortize the
overheads of the relatively slow interpreter.

\begin{figure}
  \centering
  \includegraphics{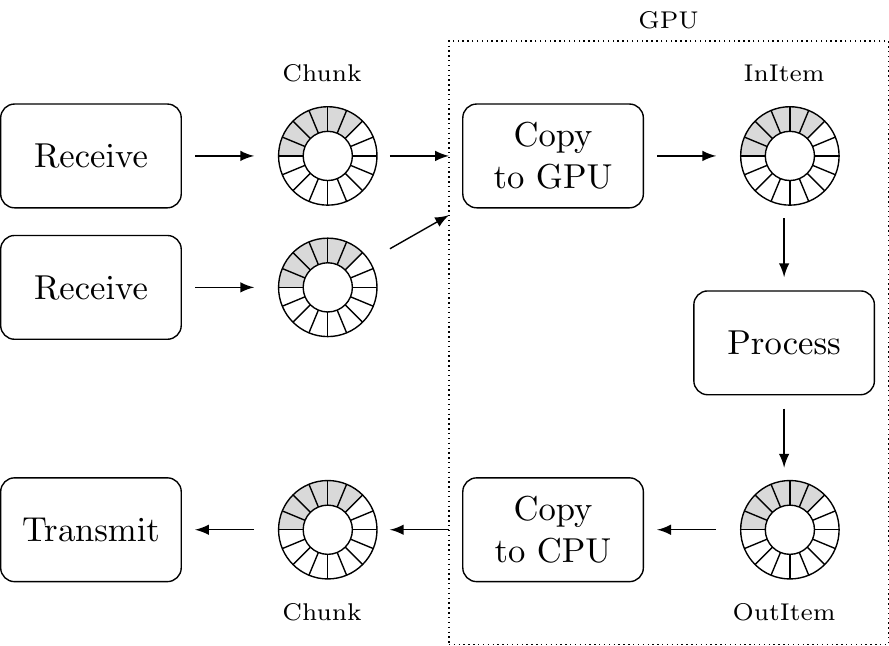}%
  \caption[Processing linked by queues]{Processing linked by queues. Where the
  diagram shows a circular buffer, the implementation uses one queue carrying
  full buffers forward and a second queue carrying empty buffers backwards.}%
  \label{fig:queues}%
\end{figure}

To ensure that all parts of the system are kept busy, we use a pipelined
design with components connected by queues of chunks, as shown in
Fig.~\ref{fig:queues}. In
particular, we need host-to-GPU transfers, GPU-to-host transfers, and GPU
computations to happen concurrently to maximize the overall throughput.
We use Python's asyncio library to manage these concurrent
operations.

\begin{figure}%
  \centering
  \includegraphics{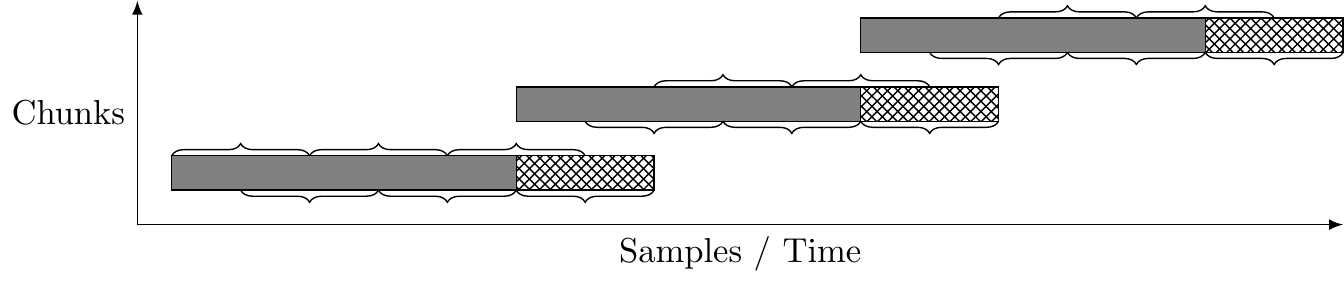}%
  \caption[Overlapping chunks]{Overlapping chunks. The cross-hatched area of
  each chunk is copied from the data in the following chunk. The braces show
  the overlapping PFB windows.}%
  \label{fig:overlap}%
\end{figure}

The PFB uses overlapping windows, which means that some
computations require data spanning an input chunk boundary. To support this,
each chunk is allocated on the GPU with some extra space at the end. The
prefix of the following chunk is then copied to this space, and computations
are performed on this expanded chunk. Provided that this extra space is at
least as large as the PFB window size, every PFB window can now be located
inside a single chunk, as shown in Fig.~\ref{fig:overlap}.

\subsection{Networking}
We use the spead2 library (a high-performance implementation of the SPEAD
protocol) both to receive input heaps from the digitizers and to transmit output
heaps to the X-engines. On the receive side, it supports collecting multiple
heaps into a chunk,
reordering them as necessary based on timestamps\cite{spead2-chunking}, before
passing the chunk to the Python code for processing. It also allows the Python
code to control the allocation of the memory: we allocate it in CUDA pinned
memory, which allows it to be efficiently copied to the GPU.

It would have been simplest to treat the two polarizations jointly in the
receive code, placing them into a single chunk. Unfortunately, design
decisions in spead2 mean that would only allow a single thread to be used, and
we were not able to achieve the required performance in this manner. Thus,
each input chunk contains only a single polarization, and Python code is used
to pair up chunks with the same timestamp. This adds complexity because this
code needs to handle corner cases where one polarization is lost.

On the transmit side, spead2 allows heaps to be defined in advance with
pointers to the data, and then transmitted many times with the values pointed
to changing each time. It also allows a list of heaps to be submitted for
transmission in one step. When allocating the output chunks, we also
create the corresponding heap structures, thus minimizing the overhead
incurred at transmission time.

\subsubsection{Data transfer}
\begin{figure}
  \centering
  \includegraphics{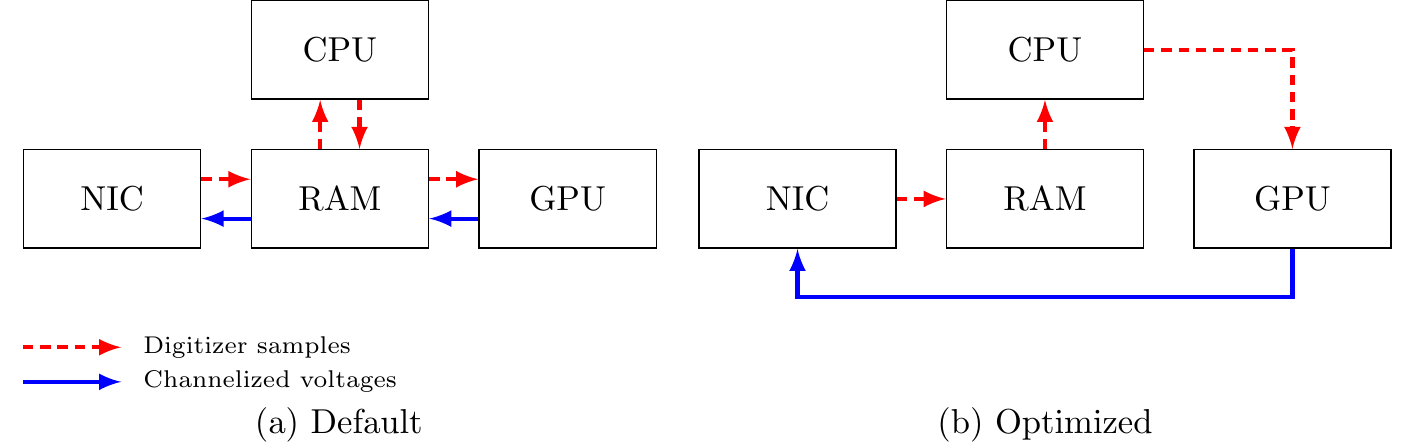}%
  \caption{System data flow}%
  \label{fig:transfer}%
\end{figure}

In the default implementation, each input sample is involved in four host
memory accesses, and each output sample in two, as shown in
Fig.~\ref{fig:transfer}(a). The NIC writes packets directly to RAM. The
spead2 library then assembles the packet payloads into chunks, again in RAM.
The final input step is that the GPU pulls the chunks from RAM. On the output
side, the GPU copies chunks to RAM, and the NIC pulls data from RAM. There is
no need for the CPU to copy data into individual packets, because the NIC is
able to gather the headers and payload for each packet from different
addresses.

NVIDIA's GPUs are able to map GPU memory into the system's address
space. This allows for a data flow that places less load on the system's RAM,
shown in Fig.~\ref{fig:transfer}(b).
Firstly, when spead2 assembles chunks, it writes directly to the GPU memory,
rather than to a staging area in host memory. Secondly, the NIC is given
pointers to GPU memory, rather than to a copy in host memory.

For the latter optimization the results are disappointing.
We found that having the NIC read directly from the GPU performs well only
when the GPU is idle; when its memory system is heavily used, the
achieved bandwidth drops to below \qty{120}{\giga\bit\per\second},
significantly less than the \qty{160}{\giga\bit\per\second} we are able to
achieve by staging through host memory.
NVIDIA recommends\cite{gpudirect} using a motherboard where the GPU and NIC
sit behind a PCIe switch, which is not the case for the systems we tested, and
so better results may be possible. It should also be noted that we were only
able to get this feature working at all on a data center GPU (A10) and not on
gaming GPUs.

Having the CPU write directly to GPU memory is more promising. We were able to
get the feature working on gaming GPUs, but with the limitation that only
\qty{256}{\mebi\byte} can be mapped. This significantly limits the maximum
chunk size, particularly when running multiple engines per GPU, and caused
performance to be lower overall.

\subsection{Coarse Delay}
The MeerKAT channelizer implements coarse delay by duplicating or removing
samples from the stream. This is easy to do with an FPGA, but less so with a
GPU since the samples are not streamed one at a time. Additionally, any PFB
windows overlapping the insertion or removal have a mix of different
coarse delays, potentially leading to artefacts. In practice the derivative of
delay is small (less than $3\times10^{-9}$ for sidereal targets) and so these
artefacts are rare.

Instead of inserting or removing samples, we handle coarse delay by
adjusting indices used to fetch samples. For each output spectrum (with a
given timestamp), we identify the appropriate position in the input stream at
which to load the data to achieve the necessary delay.
This approach allows for absolute delays that are essentially unlimited (even
negative), and every PFB window uses a consistent delay.
However, large step changes in delay (such as when
switching targets) can be problematic if they require access to older data that
has already been discarded. We can protect against this by increasing the size
of the overlap zone shown in Fig.~\ref{fig:overlap} by a number of samples
equal to the largest desired instantaneous delay change. This is not a major issue
for a dish array as a big change in delay center usually requires the dishes
to be slewed, during which time the data will be discarded anyway.

A similar problem is that the two polarizations may have different delays,
although the difference is expected to be very small since the delays are
dominated by the geometric component, which is common. Provided the overlap
zone is large enough, we can always find a pair of chunks with the same
timestamp that holds the data for both polarizations.

\subsection{Polyphase Filter Bank}\label{sec:implementation-pfb}%
In this subsection we describe only the filtering step (Eq.~\ref{eq:fir}). The
FFT step is described in the next subsection.

It is easier to implement the filter efficiently if the coarse delay can be treated
as a constant. As noted previously, coarse delay changes are rare, so we
handle this by splitting each chunk into regions with fixed coarse delay and
using a separate kernel invocation for each region. We
will thus ignore it in the following exposition, as it is simply an index
offset in the chunk. However, it should be noted that this offset cannot be
handled with pointer arithmetic, as pointers are byte-aligned while our 10-bit
samples are not.

\begin{figure}
  \centering
  \includegraphics{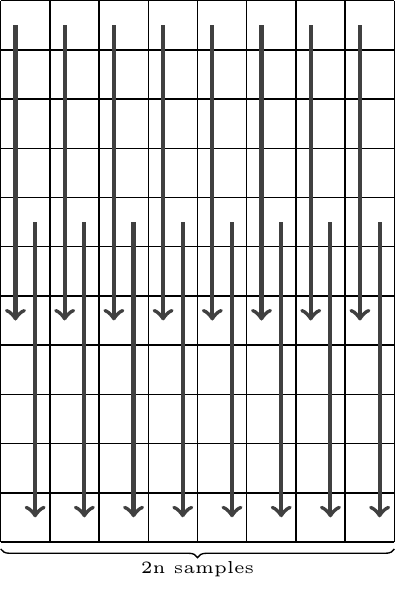}%
  \caption[Relationship of threads to input samples in the PFB]{Relationship
  of threads to input samples in the PFB. Each black arrow shows the input
  samples loaded by a single thread, for a 4-tap PFB.}%
  \label{fig:pfb-threads}%
\end{figure}

The input samples can be viewed as a 2D array with width $2n$, in which the
$i^{\text{th}}$ column undergoes a FIR filter with weights $w_i$. This maps
easily to CUDA, with one thread for each column. The thread holds $w_i$ in
its registers, as well as a sliding window of input samples, thus
minimizing the number of memory accesses required. However, this does not
provide sufficient parallelism to fully occupy the GPU: at least hundreds of
thousands of threads are needed. We thus split each column into smaller
pieces, with a thread per piece. While the output space is completely
partitioned between threads, some inputs are loaded by multiple threads, as
shown in Fig.~\ref{fig:pfb-threads}. There is thus a trade-off between having
too few threads (and not fully utilizing the GPU) and too many (and performing
many redundant loads). A heuristic we found worked reasonably well (but which
may be need to be tuned to the GPU model) is to ensure that each thread
computes at least $8T$ outputs, where $T$ is the number of taps, unless this
would lead to fewer than \num{131072} ($2^{17}$) threads.

\subsection{Fast Fourier Transform}\label{sec:implementation-fft}%
Due to coarse delays, each invocation of the PFB FIR kernel produces a
variable amount of data. We keep invoking it until we have enough data to fill
an output chunk. The last invocation may need to be truncated to avoid
overrunning the output buffer. Once this is done, we use a library to
perform a batched 1D Fast Fourier Transform.

\begin{figure}
  \centering
  \includegraphics{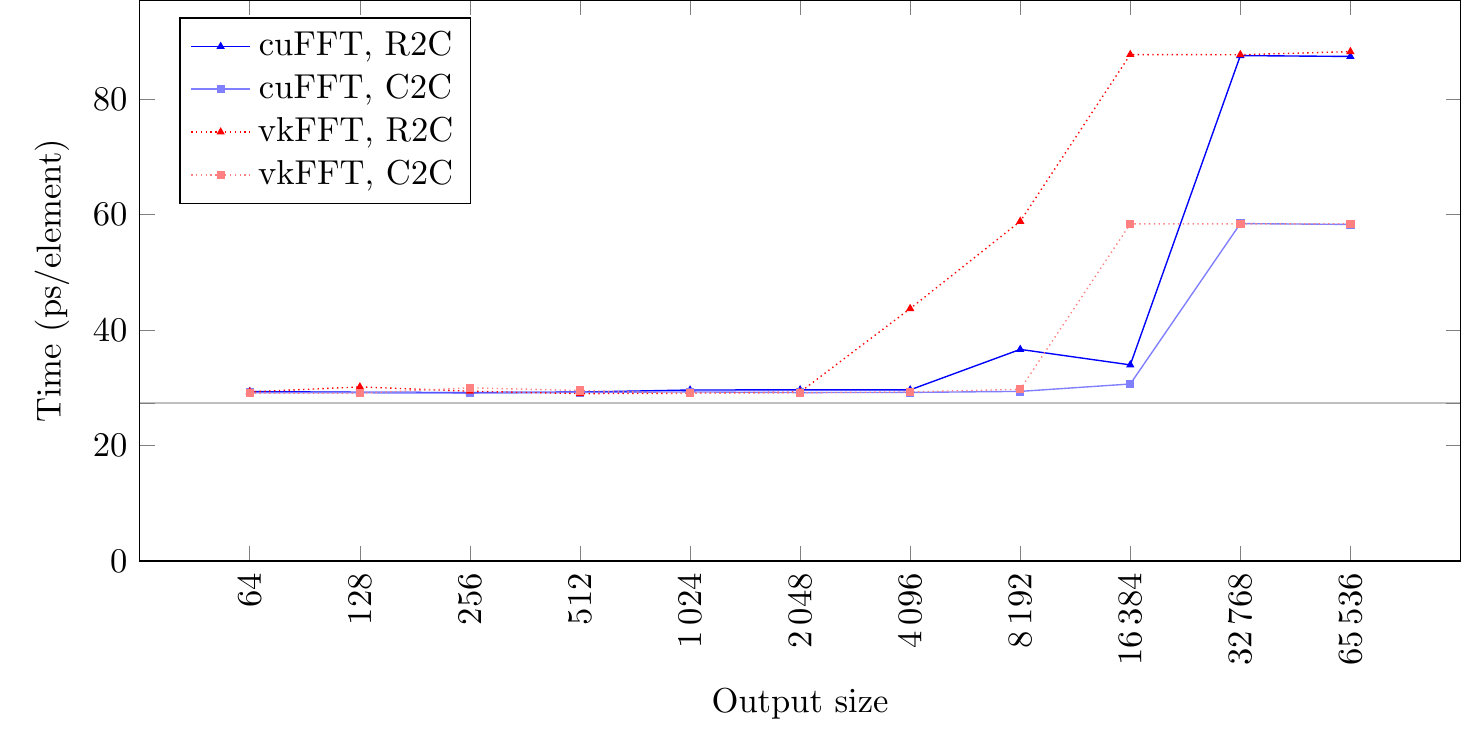}
  \caption[FFT library performance]{FFT library performance, with batched
  single-precision 1D transformations. All results use $2^{25}$ output
  values. Time is given per complex output value, and the horizontal line
  indicates the theoretical bound given by the memory bandwidth of the GeForce
  RTX~3070~Ti. Note that for real-to-complex transforms the input size is
  twice the output size.}
  \label{fig:fft}
\end{figure}

We have considered two libraries for the FFT: cuFFT
(provided as part of CUDA) and vkFFT\cite{vkfft}. The latter is highly
configurable; we have used the defaults, except that the transformation is
out-of-place. Figure~\ref{fig:fft} shows the performance of these two
libraries on real-to-complex and complex-to-complex transforms. It is
clear that for batched 1D transforms with the sizes of interest, cuFFT has
superior performance, and so we do not consider vkFFT further.

\subsubsection{FFT precision}
For MeerKAT the digitizer samples are 10-bit signed integers and the F-engine
outputs are 8-bit signed integers. Since half-precision floating point (FP16)
has a 10-bit mantissa, one might expect that a half-precision
FFT would be sufficient, as rounding errors would be smaller than the
quantization errors associated with the input and output.

\begin{figure}
  \centering
  \includegraphics{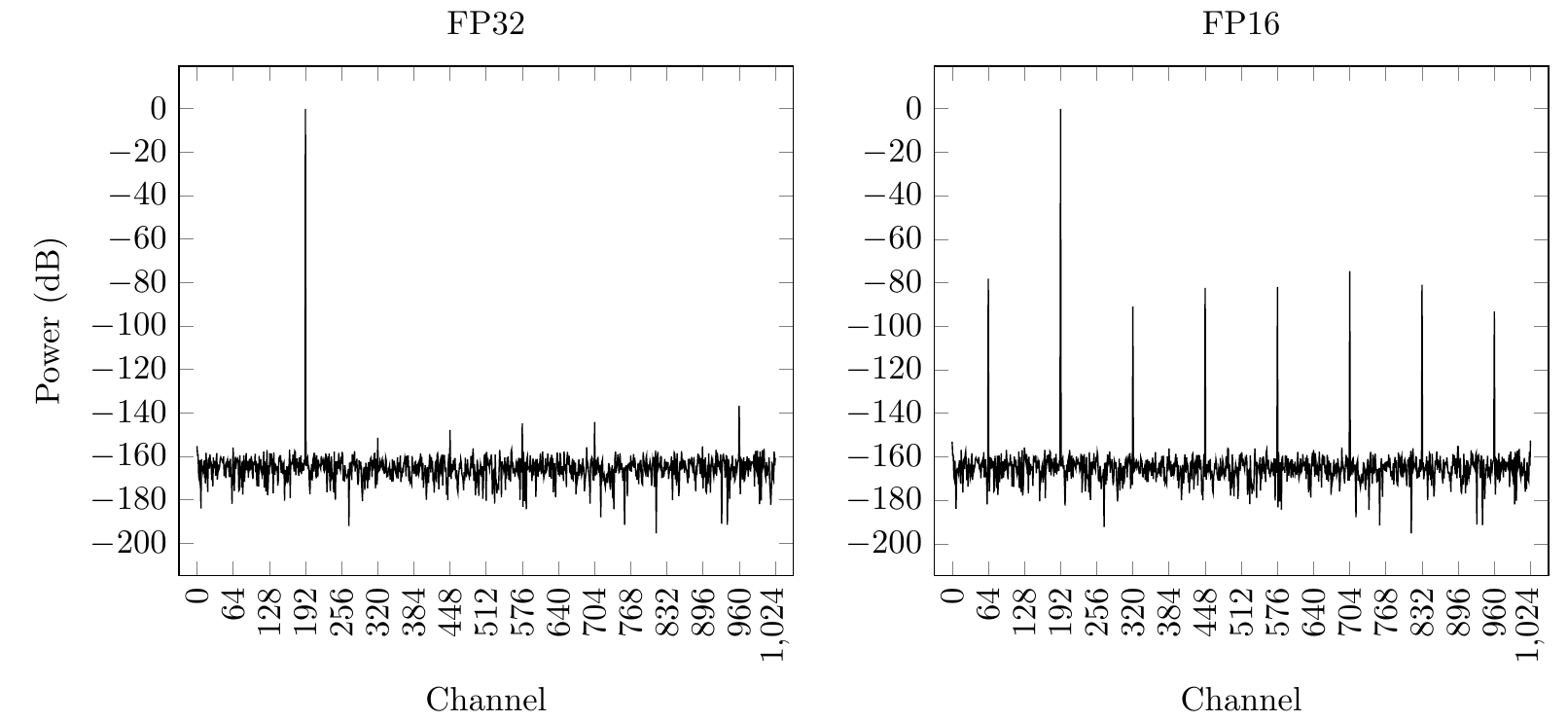}
  \caption[FFT simulation using FP32 and FP16]{FFT simulation using FP32
  (left) and FP16 (right), averaged over $2^{24}$ iterations. Only power is
  shown (not phase). The tone is in channel 192 of 1024.}%
  \label{fig:fft16}
\end{figure}

While this may be true for a single instance of the FFT, it ignores the
statistical properties of the errors. Provided the signal is suitably
dithered\cite{dither}, quantization errors will have zero mean and will
not affect the expected value of the Fourier Transform. In contrast, the
rounding errors in the FFT are data-dependent, have spectral features, and
have non-zero mean.
It is known that fixed-point PFB implementations for radio
astronomy benefit from extra internal precision for the FFT\cite{pfb-precision}
(MeerKAT uses 22-bit registers\cite{meerkat-cbf}), but
we are not aware of any studies for low-precision floating point.
To test the effect of using an FP16 FFT, we synthesized some data as follows:
\begin{enumerate}
  \item Generate a tone at a fixed frequency.
  \item Quantize it to 10-bit signed integers, using a uniform dither.
  \item Perform a real-to-complex transform using cuFFT, in either FP16 or FP32.
  \item Repeat the above many times and average the results (in double precision).
  \item Square the absolute values of the averages to convert voltage to power.
\end{enumerate}
An example of the results are shown in Fig.~\ref{fig:fft16}. While the noise
floor is similar, there are harmonics at around \qty{-75}{\dB}.
The period of the features varies depending on the binary representation of
the channel number used for the tone, reflecting the structure of the FFT.
This is significantly higher than the noise floor of the polyphase filter bank
(Fig.~\ref{fig:channel-shape}), and in the MeerKAT environment could
potentially cause strong sources of narrow-band RFI\cite{rfi-environment} to
contaminate useful parts of the band. We thus chose to stick with FP32 for the
FFT.

\subsection{Post-processing}
The remaining steps are applying fine delay, bandpass corrections, and
quantizing to 8-bit signed Gaussian integers. These are all quite
straight-forward to implement, as they can be computed independently for each
sample. We use inline PTX (CUDA's intermediate representation) to perform the
quantization with rounding and saturation.

Additionally, the data is transposed: the input is time-major,
channel-minor, but the layout expected by the X-engines is channel-major
within each heap. The transposition is done in shared memory to improve the
memory access pattern\cite{transpose}.

\subsection{Lost Data}
So far we have assumed a lossless network in which all expected packets actually
arrive. While we aim to have enough headroom that data loss does not routinely
happen, we still need to handle it gracefully. Within each chunk, spead2 sets
flags indicating which heaps were actually received. This information is
carried through the pipeline: if the window of input samples for a PFB has any
missing data, the output spectrum is flagged as unusable. Any output heap that
contains unusable spectra is simply not transmitted. This may also cause
usable spectra to be discarded, but since this is not expected to occur during
normal operation we have not attempted to optimize it.

%% file: unzip.tex
\section{Optimizing the Fourier Transform}%
\label{sect:unzip}%
In Fig.~\ref{fig:fft} it is clear that there is a large penalty for FFT
sizes above \num{16384}, and that this penalty is worse for real-to-complex
transforms. This threshold is the point at which cuFFT switches from doing the
entire transform in a single pass, to performing two (for C2C) or three (for
R2C) passes over the memory.

To eliminate this penalty for larger channel counts, we stop treating the
FFT as a black box, and split off some of the work to the other kernels.

\subsection{Real-to-Complex Transform}
We will start by eliminating the extra pass required for the real-to-complex
transform. While the cuFFT documentation does not describe how real-to-complex
transforms are implemented, the name of the final kernel
(\verb"postprocessC2C_kernelmem") suggests that it uses a technique that first
treats the even and odd elements as real and imaginary components of complex
numbers, performs a complex-to-complex transform, then performs
post-processing to get the final result\cite{real-fft}.

Instead of having cuFFT apply this technique, we can apply it manually, with
cuFFT handling just the complex-to-complex step. The advantage of doing the
post-processing ourselves is that it can be integrated into the
post-processing kernel, thus eliminating a round trip to memory.

\subsection{Unzipping the FFT}
The two passes used by cuFFT's complex-to-complex transform correspond to the
``four-step'' FFT\cite{four-step-fft}, in which a transform of size $ab$ is
decomposed into $b$ transforms of size $a$ followed by $a$ transforms of size
$b$, with the smaller transforms all computed within a faster level of the
memory hierarchy (in this case, on-chip shared memory).

As in the previous subsection, we can improve efficiency by performing this
decomposition ourselves, and merging some of the steps with existing kernels.
Our approach is actually based on the ``six-step'' FFT\cite{four-step-fft}:
\begin{enumerate}
  \item Transpose the input data, interpreted as an $a\times b$ matrix, to a
    $b\times a$ matrix (all matrices being row-major).
  \item Perform $b$ individual $a$-point FFTs.
  \item Multiply the resulting $b\times a$ matrix by appropriate roots of
    unity (the so-called ``twiddle factors'').
  \item Transpose this $b\times a$ matrix into an $a\times b$ matrix.
  \item Perform $a$ individual $b$-point FFTs.
  \item Transpose the resulting $a\times b$ matrix into a $b\times a$ matrix,
    which can be interpreted as a one-dimensional $ab$-element array.
\end{enumerate}

In our implementation, there are no explicit transposition passes; instead,
indexing of the surrounding operations is adjusted to take the transposition
into account. This makes the transposition ``free'' in the sense that it
does not directly cause extra memory transfers, but it does lead to less
efficient memory access patterns as contiguous accesses are replaced with
strided access.

We incorporate step 1 into the PFB FIR kernel (adjusting the addresses at
which values are written), perform step 2 with cuFFT, and fold the remaining
steps into the post-processing kernel.

We refer to $b$ as the ``unzipping'' factor. While the four-step FFT is normally
used with $a$ and $b$ having similar magnitude, we prefer to use a small
value, specifically $b = 4$, for several reasons:
\begin{itemize}
  \item We need to implement our own $b$-point FFT inside the post-processing
    kernel. While writing a general FFT implementation handling a range of
    sizes (even if only powers of two) is a major undertaking, a 4-point FFT
    is simple to code.
  \item Our $b$-point FFT implementation operates serially, holding all the
    data in registers of a single thread. Larger values of $b$ thus create
    more register pressure, and would probably require a rewrite using a
    parallel implementation. This is exacerbated by the post-processing for
    the real-to-complex transform, which requires two such FFTs to be
    computed by the same thread.
  \item The implicit transpositions result in access strides of $b$ elements,
    so smaller values of $b$ have better data locality.
\end{itemize}
Fig.~\ref{fig:fft} shows that the cost of step 2 is largely independent of $a$
provided it is at most \num{8192}, so the increase in $a$ that comes from
reducing $b$ is not an issue. For simplicity we have kept $b = 4$ for all
channel counts (\num{1024}--\num{32768}).

We also attempted to make the transpositions more explicit by using shared
memory\cite{transpose}, but found that the synchronization overheads
outweighed the benefits. It is possible that a more sophisticated
implementation (for example, using recent CUDA asynchronous APIs) would
achieve better results.

%% file: results.tex
\section{Results}%
\label{sect:results}%
\subsection{Hardware}%
Unless otherwise noted, all results are for a GeForce RTX 3070 Ti GPU. To improve
reproducibility, we have locked the graphics and memory clocks to
\qty{1575}{\mega\hertz} and \qty{9251}{\mega\hertz} respectively, which gives
theoretical performance of \qty{19.35}{\tera\flop\per\second} (single
precision) and \qty{592}{\giga\byte\per\second}. Despite this, we have found
that the performance of the post-processing kernel drops by 20--25\% if it is
repeated thousands of times in a tight loop, so the results for that
kernel are measured on \qty{1000} iterations at a time. This does not seem to
occur when mixed with the other kernels in real-world usage.

The CPU is an AMD EPYC 7313P (Milan) with 16 cores, \qty{3}{\giga\hertz} base
clock and \qty{3.7}{\giga\hertz} boost clock, equipped with
\qty{64}{\gibi\byte} of DDR4-3200 on a Supermicro H12SSL-i motherboard.
We considered disabling the boost clock to
give more consistent results (similar to locking the GPU clocks), but found
that doing so made a huge reduction in performance and did not substantially
improve consistency. We thus chose to keep the boost clocks enabled so that
results correspond more closely to real-world usage. Our tests are
relatively short-running, and it is possible that performance will decline
in a system that runs continuously due to thermal limitations.

The network card is an NVIDIA ConnectX-6 Dx with dual
\qty{100}{\giga\bit\per\second} ports.

\subsection{Channel Response}
\begin{figure}
  \centering
  \includegraphics{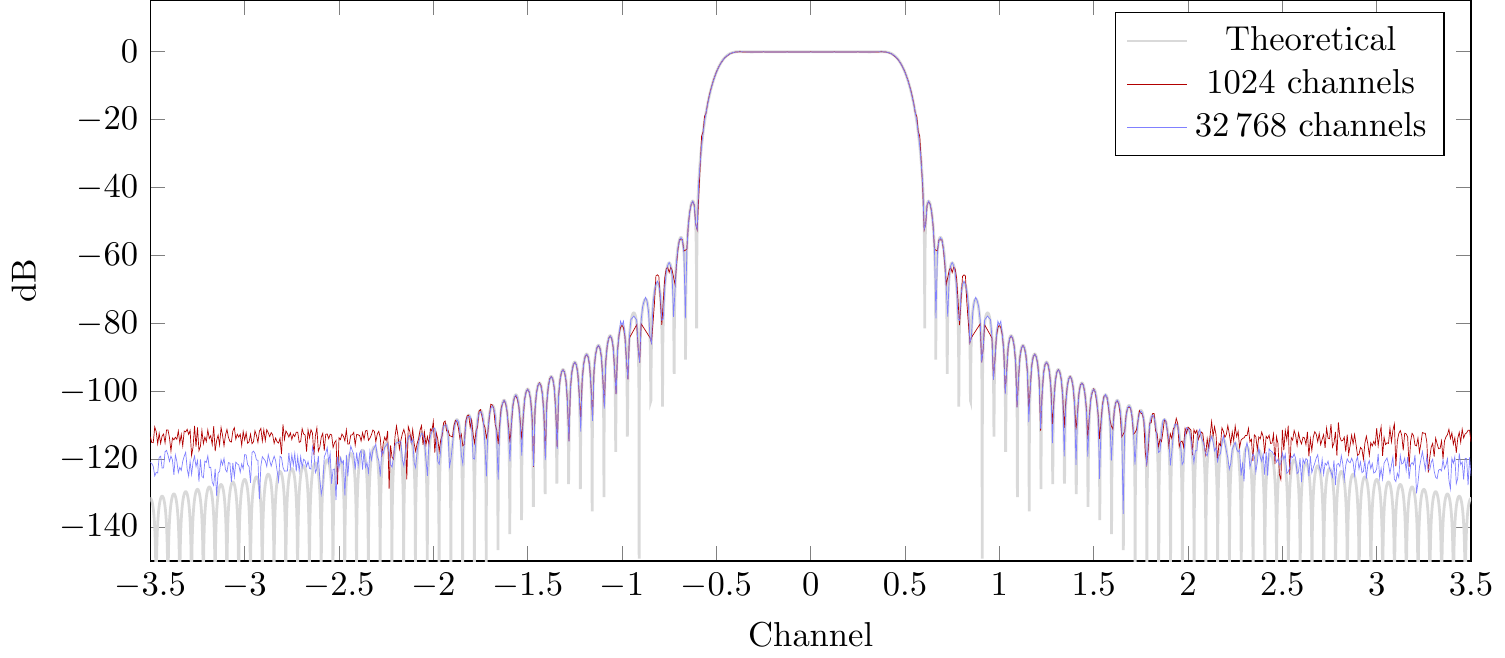}%
  \caption{Channel response for \qty{1024} and \qty{32768} channels with 16 taps}%
  \label{fig:channel-shape}%
\end{figure}

To measure the channel response, we use a simulated digitizer that generates a
common full-scale tone in both polarizations, but with independent dithering.
We then cross-correlate the F-engine outputs and integrate over
\qty{8}{\second} (we use a cross-correlation rather than an auto-correlation
so that the dithering noise is uncorrelated). By varying the frequency of the
tone by small (sub-channel) amounts, we can determine the channel response of
the engine. The 8-bit F-engine output does not have enough dynamic range to
give a full picture, so we use different gain settings for different tones.
Figure~\ref{fig:channel-shape} shows the result for \num{1024} and \num{32768}
channels and 16 taps. It also shows the theoretical ideal computed by taking
the Fourier transform of the PFB weights (a Hann-windowed sinc filter). There
is extremely good agreement down to a noise floor around \qty{-120}{\dB}. We believe the
noise floor is higher with fewer channels because the ratio of coherent
gain (gain for narrow-band signals) to incoherent gain (gain for white noise)
depends on the channel count.

\subsection{GPU Throughput}

\subsubsection{Polyphase filter-bank}
\begin{figure}
  \centering
  \includegraphics{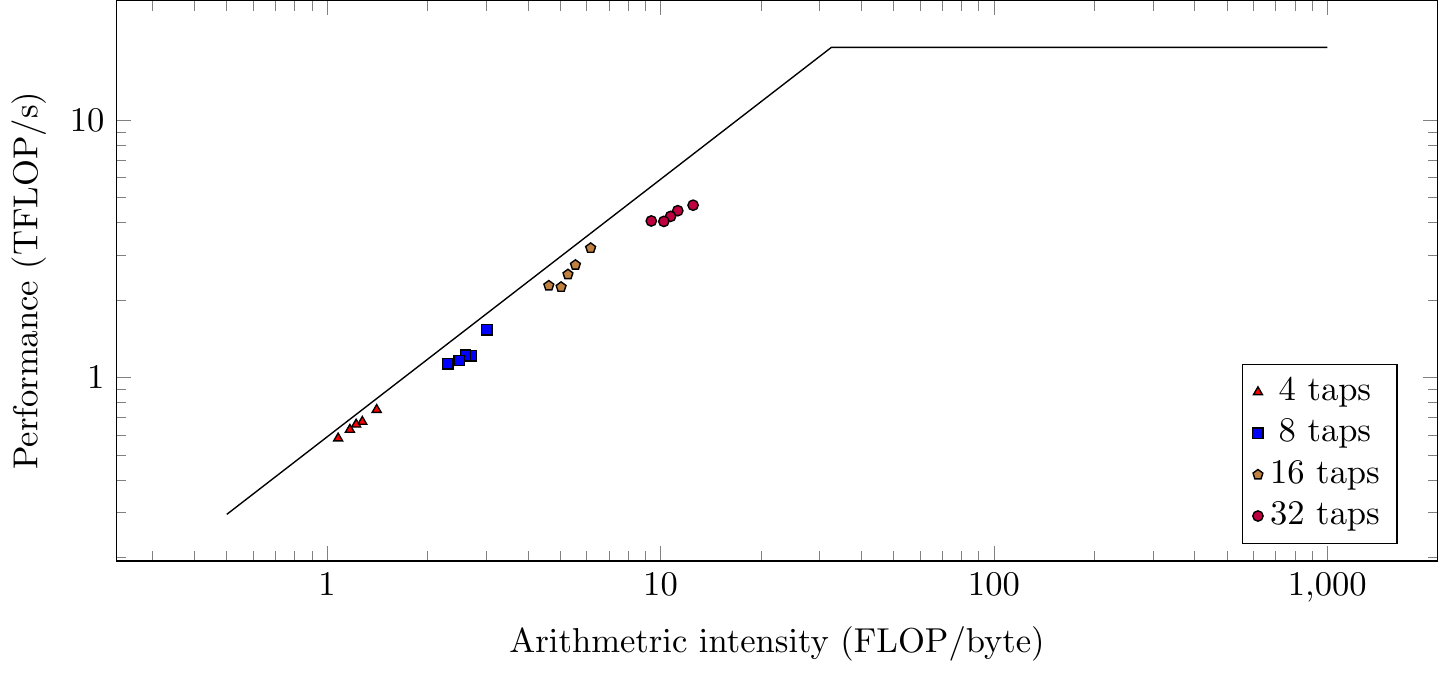}
  \caption[Roofline plot for PFB FIR filter]{Roofline plot for PFB FIR filter.
  Within each cluster, the points are for 16, 12, 10, 8 and 4 bits per sample
  from left to right. All results are for \num{32768} channels. The line
  indicates theoretical maximum performance.}
  \label{fig:roofline-pfb}
\end{figure}

Figure~\ref{fig:roofline-pfb} shows a ``roofline'' plot of the performance of
the pre-processing filter kernel, for \num{32768} channels (the results for
other channel counts are qualitatively similar). All the results are in the
left-hand side of the graph, indicating that memory accesses dominate the
performance. The configurations with up to 16 taps all use 75\% or more of the
available memory bandwidth. However, as the number of taps goes up, the
number of registers needed increases, the number of threads that can be run
concurrently decreases and the GPU's ability to hide memory latency is
reduced. With 32 taps, the theoretical occupancy (fraction of the theoretical
maximum number of concurrent threads) is 41.67\%.

\begin{figure}
  \centering
  \includegraphics{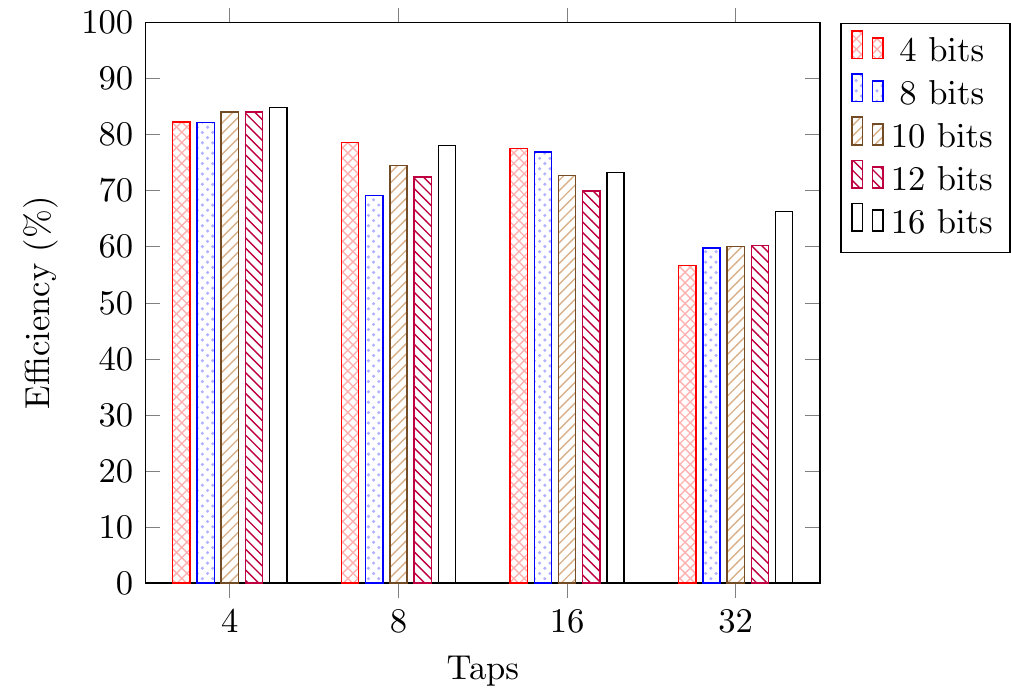}
  \caption[Efficiency of the PFB FIR filter]{Efficiency of the PFB FIR
  filter. All results are for \num{32768} channels.}
  \label{fig:efficiency-pfb}
\end{figure}
While the roofline plot shows that the memory accesses that do occur are
performed efficiently, it does not consider that some memory accesses are
redundant. The kernel loads many bytes more than once, and if this is not
absorbed by the caches, it will harm the throughput. The maximum potential
throughput of the kernel can be computed from the total size of the input
and output buffers and the theoretical bandwidth of the device.
Figure~\ref{fig:efficiency-pfb} shows the achieved efficiency relative to this
ideal, again for \num{32768} channels.

The results above all use factor-4 unzipping. This results in uncoalesced
memory writes, which reduces the performance by 4\% on average over the test
scenarios, and 19\% in the worst case.

\subsubsection{Fourier transform}
For power-of-two sizes from \num{64} up to \num{16384} (the largest size for
which cuFFT uses a single pass), the complex-to-complex FFT is memory bound:
arithmetic intensity is at most 4.5 flops per byte, and at least 87\% of the
memory bandwidth is used (both of these occur at the largest size).

\subsubsection{Post-processing}
\begin{figure}
  \centering
  \includegraphics{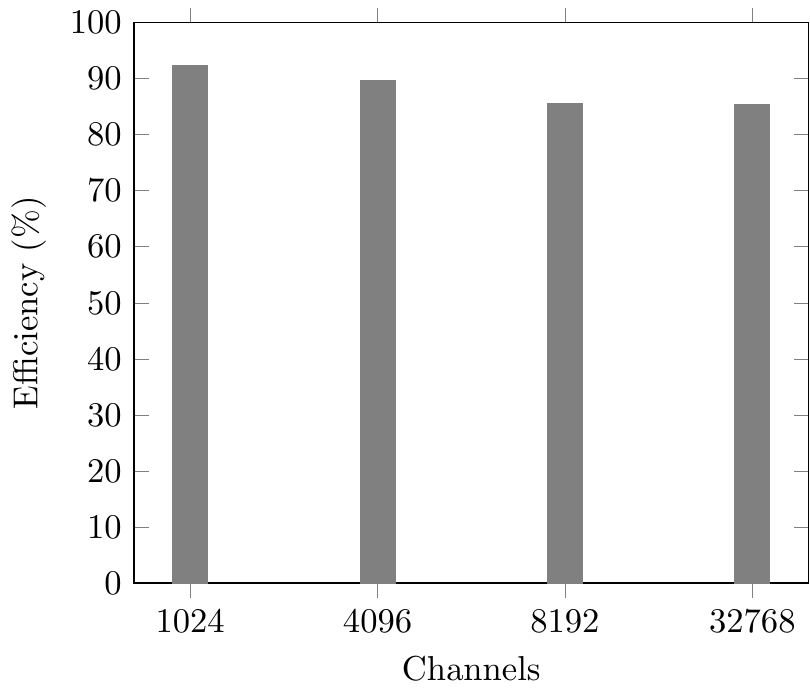}
  \caption{Efficiency of the post-processing kernel}
  \label{fig:efficiency-postproc}
\end{figure}

As with the other kernels, the post-processing is memory-bound, with an
arithmetic intensity of 6--7 flops per byte. Figure~\ref{fig:efficiency-postproc} shows the
efficiency relative to the ideal of accessing every input and output value
once at the theoretical maximum bandwidth. The efficiency declines with
increasing channel counts because the memory access pattern causes some cache
lines to be loaded multiple times. For \num{32768} channels, an extra 7\%
memory traffic is generated.

\subsubsection{Overall GPU throughput}
For even a mid-range GPU, the maximum sampling rate that can be handled is
limited by PCIe bandwidth rather than the computations on the GPU. For each
combination of PFB taps and input bit depth, we have estimated the GPU memory
bandwidth required to ensure that it does not become the bottleneck. To make
this estimate, we used the following process:
\begin{enumerate}
  \item Assume a PCIe bandwidth of \qty{160}{\giga\bit\per\second} in each
    direction (achievable on NVIDIA Ampere GPUs with a little headroom), and
    from this determine a sampling rate.
  \item Measure the time required to run all the kernels on our test system,
    and use linear scaling to determine a memory bandwidth of a hypothetical
    GPU that would run the kernels just fast enough.
  \item Since the measurement above does not include any PCIe transfers, add
    the GPU memory bandwidth required for transferring the inputs and outputs
    over PCIe. Similarly, add time to copy the prefix of each chunk to the
    suffix of the previous chunk. In both cases we assume 100\% efficiency
    in the memory accesses.
\end{enumerate}

\begin{figure}
  \centering
  \includegraphics{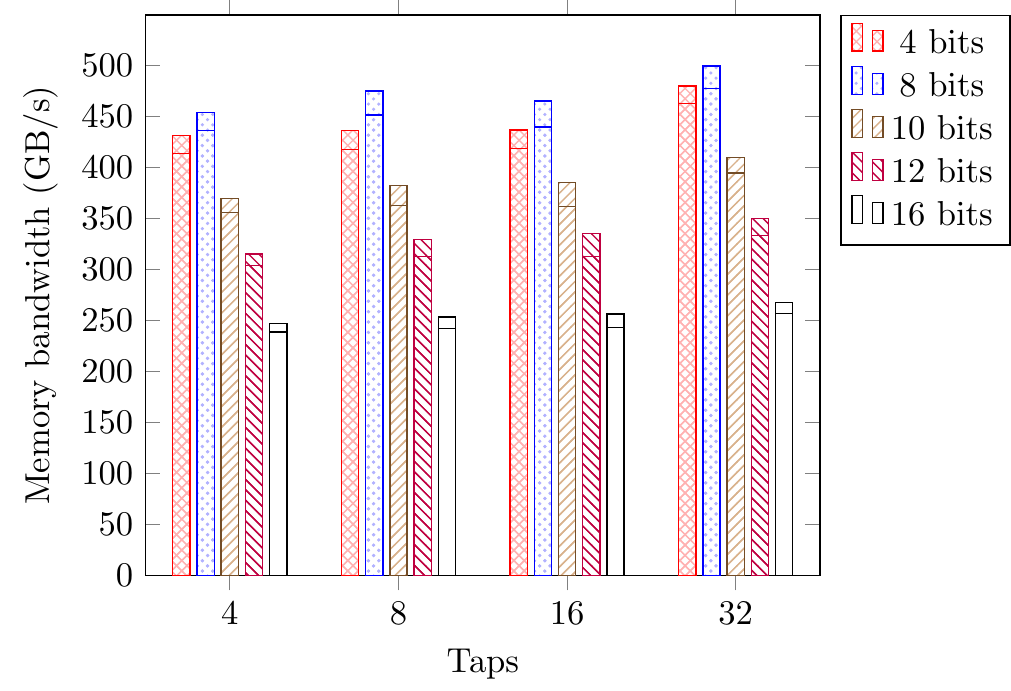}
  \caption[Estimated GPU memory bandwidth]{%
    Estimated GPU memory bandwidth. This is the minimum bandwidth the GPU will
    require for the computations to become bottlenecked by a PCIe 4 bus.
    The top of each bar represents \num{32768} channels while the horizontal
    line near the top represents \num{1024} channels.
  }
  \label{fig:compute-bw}
\end{figure}

Figure~\ref{fig:compute-bw} shows the results for \num{1024} and \num{32768}
channels. It may seem counter-intuitive that going from 8 to 16 bits
per sample reduces the required bandwidth. This occurs because the PCIe
bandwidth is kept constant and hence the sampling rate decreases. The bulk of
the on-GPU memory traffic is performed in single-precision floating point
rather than scaling with the input bit depth, and so decreasing the sample
rate decreases the memory bandwidth needed for that traffic.

To validate this model, we can artificially limit the memory clock on our test
hardware to simulate a lower-end GPU. This is not a perfect test, since a
lower-end GPU would generally have fewer streaming multiprocessors and a
narrower memory bus, but should give some indication of the accuracy. Our GPU
only supports a few fixed memory frequencies, so we fix it to
\qty{810}{\mega\hertz}, which gives a theoretical bandwidth of
\qty{51.84}{\giga\byte\per\second}. At \num{32768} channels, 10-bit samples,
and 16 taps, the model indicates a maximum sampling rate of
\qty{1077}{\mega\sample\per\second}. In practice, we found
\qty{930}{\mega\sample\per\second} was the highest rate we could run the
full engine without falling behind the incoming data (to the nearest
\qty{10}{\mega\sample\per\second}). This shows that there are additional
overheads not accounted for by the model, but (at least for this case) they
are less than 15\%.

\subsection{System Tuning}\label{sec:results-system}%
We found that we needed to do a substantial amount of system-level tuning to
obtain good performance, using a combination of hardware placement, BIOS
settings and kernel settings.

To test the throughput of the whole system, we run either one or four
instances of the F-engine on the system under test. Input data is
provided by digitizer simulators (one per F-engine) running on another
machine. The output data is sent into the network (as multicast streams).
Using four engines is representative of how the code is expected to be
deployed for the MeerKAT Extension, where the highest sampling rate will be
\qty{1750}{\mega\sample\per\second}. In this case, each engine is assigned to
a single quadrant of the CPU, and hence to a single Core Complex Die (CCD).
Tests with a single engine are aimed at measuring the maximum bandwidth
achievable with the current implementation, and use one thread per CCD (one
network receive thread per polarization, one network transmit thread and the
main Python thread).

Since the UDP protocol in use is lossy, it is not possible to measure maximum
throughput directly. Rather, we repeat a number of experiments in which a
fixed sampling rate is chosen, and we observe the F-engine input over
\qty{20}{\second} to check if there are any gaps in the received timestamps
(indicating packet loss). The engine is allowed to run for a few seconds
before this observation period begins, as it is quite common for some packets
to be lost while the process ``warms up''. We then use a binary search to
determine the highest sampling rate for which no packets are lost, to the
nearest \qty{10}{\mega\sample\per\second}. As the sampling rate approaches the
critical rate at which the implementation can keep up, packet loss during the
\qty{20}{\second} window becomes a random event, and so we see variation of a
few percent even when the configuration is not changed.

These results should be seen as upper bounds, as running for \qty{20}{\second}
under ideal conditions (for example, with no changes to delay) does not
guarantee stable operation in real-world use.

\subsubsection{BIOS settings}
Table~\ref{tbl:downgrade-bios} shows achieved sampling rates in each case,
starting with our optimized system as a baseline and then showing the impact of
changing one setting at a time (except for the row marked ``BIOS defaults'').
These results all use \num{32768} channels,
16-tap PFBs and 10-bit digitizer samples (a representative configuration for
the MeerKAT Extension). The chunk size is $2^{26}$ samples for one
engine or $2^{24}$ samples per engine when using four engines.

\begin{table}
  \centering%
  \caption[Effect of BIOS settings on sampling rate]{%
    Effect of BIOS settings on sampling rate. Values marked with a *
    are the effective BIOS defaults (the nominal default in most cases is
    ``Auto''). Percentages are relative to the baseline configuration.}%
  \label{tbl:downgrade-bios}%
  \begin{filecontents*}{data/downgrade/bios.csv}
setting,baseline,test,default,n1,n4,Notes
Baseline,,,,6260,2080,
BIOS defaults,,,,4550,1610,
APBDIS,1,Auto*,Auto,6240,2080,Strangely no loss – maybe in+out gives enough load?
DF Cstates,Disabled,Enabled*,Auto (Enabled),5740,2070,
LCLK Frequency Control,593 MHz,Auto*,Auto,6220,2070,
NUMA nodes per socket,NPS1*,NPS2,Auto (NPS1),4970,1970,
NUMA nodes per socket,NPS1*,NPS4,Auto (NPS1),2780,1610,
Preferred IO bus,GPU,Auto*,Auto,6210,1880,
Preferred IO bus,GPU,NIC,Auto,6030,1880,
PCIe relaxed ordering,Enabled*,Disabled,Enabled,5780,1150,
Local APIC mode,x2APIC,Auto*,Auto,6170,2090,
IOMMU,Disabled,Enabled*,Auto (Enabled),6240,1770,
\end{filecontents*}
\pgfplotstabletypeset[
    every head row/.style={
      before row={%
        \toprule
        &&&\multicolumn{2}{c}{1 engine} & \multicolumn{2}{c}{4 engines}\\
        \cmidrule(lr){4-5}%
        \cmidrule(lr){6-7}%
      },
      after row={\midrule},
    },
    every last row/.style={
      after row={\bottomrule},
    },
    col sep=comma,
    columns={setting,baseline,test,n1,n1p,n4,n4p},
    columns/setting/.style={
      column name=Setting,
      string type,
      column type=l,
    },
    columns/baseline/.style={
      column name=Baseline,
      string type,
      column type=c,
    },
    columns/test/.style={
      column name=Tested,
      string type,
      column type=c,
    },
    columns/n1/.style={
      column name={\si{MS/s}},
    },
    columns/n4/.style={
      column name={\si{MS/s}},
    },
    columns/n1p/.style={
      column name=\%,
      fixed zerofill,
      precision=1,
      dcolumn={D{.}{.}{-1}}{c},
    },
    columns/n4p/.style={
      column name=\%,
      fixed zerofill,
      precision=1,
      dcolumn={D{.}{.}{-1}}{c},
    },
    create on use/n1p/.style={
      create col/expr={\thisrow{n1} / 62.60},
    },
    create on use/n4p/.style={
      create col/expr={\thisrow{n4} / 20.80},
    },
    section start/.style={before row={\midrule}},
    every row no 2/.style={section start},
    every row no 5/.style={section start},
    every row no 10/.style={section start},
  ]{data/downgrade/bios.csv}%
\end{table}

The BIOS settings chosen are a combination of those recommended by
AMD\cite{epyc-workload-tuning,epyc-hpc-tuning} and our own experience and
experimentation. Not all of the settings recommended by AMD are available on
our motherboard.

The first group of settings (starting with APBDIS) relate to power management.
Because the F-engine operates on a batch of data then becomes idle until the next
batch is ready, it may cause some part of the system to drop into a
lower-power, less-performant state. There is usually a latency to return to
full performance, and if that is too high it can lead to data loss. In this
case it appears that only DF Cstates are beneficial. In smaller
microbenchmarks we have seen APBDIS cause poor performance at certain data
rates---usually not the highest data rates, but rather ones that are low
enough to allow the low-power state to be engaged.

CPU power management (P-states and C-states) is also important, but we chose
to control that through the operating system rather than the BIOS.

The next group relate to the way memory accesses are performed. By default,
memory addresses are interleaved across all the memory channels (1 NUMA node
per socket, or NPS1), but the memory channels can also be partitioned into two
or four sets (NPS2/NPS4) where the OS can allocate memory from specific sets.
This can reduce memory latency if the users of the memory (CPU cores or
PCIe devices) are located close to the memory controllers. A single PCIe bus
can also be designated as the ``preferred IO bus'', and will get priority when
there is contention for memory access. Finally, ``PCIe relaxed ordering''
allows PCIe transactions to proceed out-of-order under some circumstances,
which can improve utilization by preventing head-of-line blocking.

We expected NPS1 to produce sub-optimal results for 1 engine, because the load
is not evenly balanced across the system. However, we expected NPS4 to work
well for 4 engines, because each engine runs on one CCD and uses the nearest
memory channels, which should give ideal load balancing and minimal latency.
Our hypothesis is that the copy engines on the GPU perform coarse-grained
time-sharing between the processes, and hence only utilise half or a quarter
of the memory channels at a time rather than having in-flight transactions on
them all concurrently. This does match AMD's recommendation that workloads
requiring accelerator throughput should use NPS1.

We were surprised that setting the GPU as the preferred I/O device was
optimal. It is commonly recommended that the NIC is the preferred I/O
device, because it has real-time requirements and will drop packets if it is
not able to transfer them quickly. However, because the whole system is
real-time, low GPU throughput can also lead to packet loss, and early
experiments suggest that the GPU does not cope well with contention for memory
access.

The final set of options concern features that can be enabled. Using x2APIC
may reduce interrupt latency, but this does not appear to be important, and
differences may be just noise. Using the IOMMU seems to reduce performance.

\subsubsection{Kernel settings}
Table~\ref{tbl:downgrade-kernel} shows the effect of kernel settings,
similarly to Table~\ref{tbl:downgrade-bios}. The first two settings control
CPU frequency scaling and low-power CPU states, and can also be controlled via
the BIOS. The latter appears not to affect performance, presumably indicating
that the full workload is sufficiently intense to prevent the CPU from
entering these deep C-states. As with APBDIS and DF Cstates, it is possible
that lower-bandwidth workloads will actually perform worse if they are light
enough to allow these low-power states to be used.

By default, the NVIDIA NIC loops outgoing IP multicast traffic into the
receive path so that processes running on the same machine can receive the
traffic. While convenient, this creates a significant overhead in transmitting
multicast data. Disabling this behavior (by writing to
\verb"/sys/class/net/*/settings/force_local_lb_disable") improves performance
with four engines. The loopback behavior is automatically disabled if there is
only one process using ibverbs, which is why the single-engine case is
unaffected. With other ibverbs processes present (but stopped, and hence using
no CPU time) the rate is reduced to \qty{5580}{\mega\sample\per\second}.

The last four options aim to reduce CPU overhead and allow the code to run
more efficiently. In the 4-engine case we are not CPU-bound, which is why they
make no difference. The \texttt{vm.nr_overcommit_hugepages} setting allows
spead2 to allocate its buffers in huge pages, which can reduce the number of
translation look-aside buffer (TLB) misses.

NUMA balancing is a kernel
mechanism which monitors which cores are using which memory
pages\cite{epyc-hpc-tuning}; it is
implemented by periodically unmapping some pages, causing the next access to
page fault. While the results show no effect, we have found in
longer tests that these page faults can cause occasional high latency leading
to data loss.

We enable real-time scheduling for the processes to ensure that they get CPU
time whenever they need it. By default, Linux does not allow real-time
processes to use more than \qty{0.95}{\second} of every second, to prevent a
malfunctioning real-time process from locking up a system. We increase this to
\qty{0.999}{\second} to allow the processes more CPU time without completely
removing the protection. Surprisingly, this makes much more than a 4.9\%
difference. We hypothesize that this is because a process that uses less than
95\% CPU on average may still exceed it during some second, and when it does
so, the \qty{50}{\milli\second} it is stalled is long enough for the network
buffer to be overrun. On the other hand, a \qty{1}{\milli\second} stall is
short enough that a process with less than 99.9\% average usage can recover
from it.


\begin{table}
  \centering%
  \caption[Effect of kernel settings on sampling rate]{%
    Effect of kernels settings on sampling rate. The tested
    values are the kernel defaults. Percentages are relative to
    the baseline configuration.}%
  \label{tbl:downgrade-kernel}%
  \begin{filecontents*}{data/downgrade/kernel.csv}
setting,baseline,test,n1,n4,Notes
Baseline,,,6260,2080,
CPU scaling governor,performance,ondemand,3030,2070,
cmdline: processor.max_cstate,1,--,6230,2080,
NIC force_local_lb_disable,1,0,6240,1400,Single-process not affected because loopback doesn't kick in without multiple processes?
vm.nr_overcommit_hugepages,2048,0,5960,2080,
kernel.numa_balancing,0,1,6260,2080,Random failures rather than general degradation
kernel.sched_rt_runtime_us,999000,950000,5210,2080,
cmdline: mitigations,off,--,5860,2070,
\end{filecontents*}
\pgfplotstabletypeset[
    every head row/.style={
      before row={%
        \toprule
        &&&\multicolumn{2}{c}{1 engine} & \multicolumn{2}{c}{4 engines}\\
        \cmidrule(lr){4-5}%
        \cmidrule(lr){6-7}%
      },
      after row={\midrule},
    },
    every last row/.style={
      after row={\bottomrule},
    },
    col sep=comma,
    columns={setting,baseline,test,n1,n1p,n4,n4p},
    columns/setting/.style={
      column name=Setting,
      string type,
      column type=l,
    },
    columns/baseline/.style={
      column name=Baseline,
      string type,
      column type=c,
    },
    columns/test/.style={
      column name=Tested,
      string type,
      column type=c,
    },
    columns/n1/.style={
      column name={\si{MS/s}},
    },
    columns/n4/.style={
      column name={\si{MS/s}},
    },
    columns/n1p/.style={
      column name=\%,
      fixed zerofill,
      precision=1,
      dcolumn={D{.}{.}{-1}}{c},
    },
    columns/n4p/.style={
      column name=\%,
      fixed zerofill,
      precision=1,
      dcolumn={D{.}{.}{-1}}{c},
    },
    create on use/n1p/.style={
      create col/expr={\thisrow{n1} / 62.60},
    },
    create on use/n4p/.style={
      create col/expr={\thisrow{n4} / 20.80},
    },
  ]{data/downgrade/kernel.csv}%
\end{table}

\subsubsection{Hardware placement}
The motherboard has five x16 PCIe 4.0 slots, but performance-wise they are not all
the same. The I/O die of the CPU is split into four quadrants. Each quadrant
supports 32 PCIe lanes, but we found that a quadrant is not able to sustain
full-bandwidth transfers between these lanes and system memory: the maximum is
around \qty{36}{\giga\byte\per\second} in each direction. We thus found
extremely poor performance when placing the GPU and the network interface card
(NIC) in a pair of slots connected to the same quadrant.

\begin{figure}
  \centering
  \includegraphics{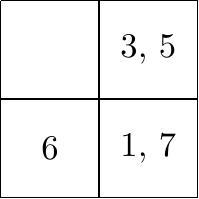}%
  \caption{Mapping of x16 PCIe slots to CPU quadrants on H12SSL-i motherboard}%
  \label{fig:slots}%
\end{figure}

Even when using slots attached to different quadrants, not all combinations
are equal. Table~\ref{tbl:slots} shows the results with various combinations
of slots, and Fig.~\ref{fig:slots} shows the association of the slots to the
quadrants of the CPU\cite{H12SSL}.
\begin{table}
  \centering
  \caption[Effect of PCIe slots on performance]{%
    Effect of PCIe slots on performance. Percentages are relative to
    the top row, which is the baseline used in other results.}%
  \label{tbl:slots}%
  \begin{filecontents*}{data/downgrade/slots.csv}
GPU slot,NIC slot,n1,n4,Notes
7,3,6260,2080,Baseline
7,1,610,150,
6,3,6170,2000,
6,1,610,150,
3,7,6180,2080,
3,5,3550,900,
\end{filecontents*}
\pgfplotstabletypeset[
    every head row/.style={
      before row={%
        \toprule
        &&\multicolumn{2}{c}{1 engine} & \multicolumn{2}{c}{4 engines}\\
        \cmidrule(lr){3-4}%
        \cmidrule(lr){5-6}%
      },
      after row={\midrule},
    },
    every last row/.style={
      after row={\bottomrule},
    },
    col sep=comma,
    columns={NIC slot,GPU slot,n1,n1p,n4,n4p},
    columns/n1/.style={
      column name={\si{MS/s}},
    },
    columns/n4/.style={
      column name={\si{MS/s}},
    },
    columns/n1p/.style={
      column name=\%,
      fixed zerofill,
      precision=1,
      dcolumn={D{.}{.}{-1}}{c},
    },
    columns/n4p/.style={
      column name=\%,
      fixed zerofill,
      precision=1,
      dcolumn={D{.}{.}{-1}}{c},
    },
    create on use/n1p/.style={
      create col/expr={\thisrow{n1} / 62.60},
    },
    create on use/n4p/.style={
      create col/expr={\thisrow{n4} / 20.80},
    },
  ]{data/downgrade/slots.csv}%
\end{table}

\subsection{Power Consumption}
We used the 4-engine test case to measure power consumption, as it places
greater demand on the system (by virtue of having greater total bandwidth). We
used a sampling rate of \qty{2000}{\mega\sample\per\second}, and other
parameters are the same as for the system tuning results.

With the GPU clocks locked to the base values, the power consumption for the
GPU, as reported by \texttt{nvidia-smi}, is \qty{156}{\watt} and the power for
the whole system, as reported by the baseboard management controller (BMC) is
\qty{407}{\watt}. Unlocking the clocks causes power usage to increase by
\qty{73}{\watt}. On the other hand, the graphics clock can be reduced as low as
\qty{660}{\mega\hertz} without causing any loss of data, but doing so saves only
\qty{8}{\watt} compared to using the base clocks.

When using a sampling rate of \qty{1712}{\mega\sample\per\second}, \num{4096}
channels (a common configuration for MeerKAT), and base GPU clocks, the system
power usage is \qty{388}{\watt}, or \qty{97}{\watt} per antenna. Within the
margins of error, this is the same per-antenna power consumption as the
current SKARAB (FPGA) platform used in MeerKAT. It should be noted that the
SKARAB platform is almost a decade old and hence is not representative of the
power consumption of more modern FPGAs.

%% file: conclusions.tex
\section{Conclusions and future work}%
\label{sect:conclusions}%
We have built a wide-band channelizer that is able to process the data for
four MeerKAT antennas on a single commodity GPU, and which implements all the
features of the existing FPGA-based wide-band channelizer. The throughput is
limited by the PCIe bandwidth of the GPU. The main outstanding work to make it
ready for the MeerKAT Extension correlator is the addition of a narrow-band
mode. We have done some prototyping of a low-pass filter kernel, and are
confident that it will be possible to implement concurrent wide-band and
narrow-band modes within the same pipeline.

The computations on the GPU are not a bottleneck for our chosen GPU (RTX 3070
Ti). Furthermore, there are GPUs available with significantly higher memory
bandwidth, so given sufficient budget, we do not expect them to become a
bottleneck for any use cases. We have thus not tried to squeeze out all the
possible performance. Nevertheless, there may be value in further
optimizations to allow cheaper and less power-hungry GPUs to be used.
Here are a number of high-level optimizations we have considered:
\begin{itemize}
  \item We have split off a small part of the FFT into the other kernels, but
    perhaps it can be completely fused, with step 2 of the six-step FFT merged into
    the PFB FIR kernel. The challenge here
    is that performing an FFT pass requires a fairly specific mapping of data
    to threads and thread blocks, which might not be compatible with the
    mapping currently used by those kernels. For channel counts that are low
    enough to support a single-pass FFT, it may be possible to fuse all three
    kernels together.
  \item While we have noted issues with computing the FFT in FP16, it may be
    feasible to use FP16 for the inputs and/or outputs of the FFT, with the
    internal computations done in FP32. The PFB FIR could then also
    potentially be performed in FP16, which would reduce register pressure and
    allow more taps to be used at the same throughput.
\end{itemize}

A common theme in these optimizations, as well as the optimizations we have
already implemented, is that modular design does not work well for
memory-bound applications. For example, we started by treating the PFB FIR,
the FFT and the post-processing as three independent modules, but to improve
performance we had to redistribute functionality between them, causing tight
coupling. Similarly, the PFB FIR kernel and the post-processing kernel are
tightly coupled to the input and output data formats, and cannot be used as-is
for a system that expects
different formats. This makes it difficult to create optimal yet
reusable code that can be mixed and matched in stream processing frameworks
such as Bifrost\cite{bifrost}, which use GPU memory as an interface boundary.

Because the implementation was originally designed for MeerKAT, it did not
target higher data rates per antenna. While ingress rates exceeding
\qty{120}{\giga\bit\per\second} are possible, they are limited by the
single-core performance of spead2. This is not a fundamental limitation, as
the network receive could be distributed across several threads, or spead2
could be replaced by a bespoke library tailored to the exact packet layout. We
expect that input rates of \qty{160}{\giga\bit\per\second} could be
achieved, as they are for the multi-engine case.

The results presented all use a single GPU. We have also experimented
with using two GPUs and two NICs per server (on a different server).
Unfortunately, performance does not scale linearly, because the system memory
bandwidth becomes a bottleneck, and we are forced to use sub-optimal PCIe
slots. Recently released CPUs may help with bandwidth: EPYC 9004-series processors double
the PCIe bandwidth (with PCIe 5.0) but more than double the memory bandwidth
(from \qty{205}{\giga\byte\per\second} to
\qty{461}{\giga\byte\per\second})\cite{epyc-genoa-datasheet},
while Xeon Max CPUs have on-board high bandwidth memory (HBM)\cite{xeon-max}.